# Polymer amide as a carrier of $^{15}$N in Allende and Acfer 086 meteorites

Malcolm. W. McGEOCH[1], Tomáš ŠAMOŘIL[2], David ZAPOTOK[3]
and Julie E. M. McGEOCH[4*].

[1] PLEX Corporation, 275 Martine St., Suite 100, Fall River, MA 02723, USA.
[2] TESCAN ORSAY HOLDINGS s.a., Libusina trida 21, 623 00 Brno, Czech Republic.
[3] TESCAN USA, Inc., 765 Commonwealth Drive, Suite 101, Warrendale, PA 15086.
[4] Dept. of Molecular and Cellular Biology, Harvard University, 52 Oxford St., Cambridge MA 02138, USA.
*Corresponding author. E-mail: mcgeoch@fas.harvard.edu

**Abstract**
Polymers of amino acids have been found in Allende and Acfer 086, with extra-terrestrial origin evidenced by isotopically enhanced satellite peaks in the 800 – 1200 Dalton mass range. The present work employs focused ion beam milling of micron scale powder to generate negative ion spectra containing CN ions at masses 26 and 27 for $^{15}$N determination. Respectively, for CN ions, Allende and Acfer 086 show $\delta^{15}$N = 410 ± 220 $^0/_{00}$ and 1,015 ± 220 $^0/_{00}$ in the powder samples. We measure an exact 1:2 ratio of $C_2$ to CN negative ions that is consistent with ion beam fragmentation of a polymer amide –CCN- backbone into diatomic fragments. It is inferred that polymer amide comprises a large part of the $^{15}$N bearing organic material in this CV3 meteorite class.

## INTRODUCTION

Meteorites have been known to have $^{15}$N (heavy nitrogen) enrichment above terrestrial since the work of (Injerd and Kaplan 1974) and (Kung and Clayton 1978). Studies have mostly been on carbonaceous chondrites, within which the highest $^{15}$N enhancements have been in the least metamorphosed and aqueously altered with extreme examples of "whole rock" $\delta^{15}$N in parts per mil being Renazzo (170$^0/_{00}$) (Kung and Clayton 1978), Bencubbin (973$^0/_{00}$) (Prombo and Clayton 1985), Bells (335$^0/_{00}$) (Kerridge 1985) and ALH 85085 (858$^0/_{00}$) (Grady and Pillinger 1990). The per mil enhancement is defined by $\delta^{15}N(\%o) = \left(\frac{\left(^{15}N/^{14}N\right)_{SAMPLE}}{\left(^{15}N/^{14}N\right)_{AIR}} - 1\right) \times 1000$ in which air has $^{15}$N/$^{14}$N = 3.612x10$^{-3}$ (IAEA Vienna 1995). The convention in astronomy is to quote isotope ratios as the inverse of this ratio, i.e. ($^{14}$N/$^{15}$N)$_{AIR}$ = 277.

Classical methods of isotope analysis by stepped pyrolysis and mass spectrometry have shown (Grady and Pillinger 1990; Lewis et al. 1983) that typically there can be chemical components with normal or even strongly depleted $^{15}$N, relative to terrestrial, and other components with a large $^{15}$N excess. An example of this is Allende, of type CV3, that exhibits less than terrestrial $^{15}$N ($\delta^{15}$N = -18.6 $^0/_{00}$) when measured in bulk (Pearson et al. 2006, and references therein), yet has a minority chemical component with potentially very high $^{15}$N that is released above 600C (Lewis et al. 1983). Another example is Murchison (class CM2), which also has a high temperature heavy component (Lewis et al. 1983), displays bulk $^{15}$N of 52.7 $^0/_{00}$ (Pearson et al. 2006), yet has chemically identified sub-components with $^{15}$N as high as 104 $^0/_{00}$ (Pizzarello et al. 1994).

As to the nature of the $^{15}$N carrier(s), (Epstein et al. 1987) reported strong $^{15}$N and D enhancements in a well-characterized amino acid fraction from Murchison (Cronin and Pizzarello 1986). The per mil enhancements for $^{15}$N (90 $^0/_{00}$) and D (1370 $^0/_{00}$) were considered to be minimum values in view of possible contaminants, or of the partial correction for hydrogen isotope exchange that had been applied. Prior to this work, (Becker et al. 1982) had measured $\delta^{15}$N = 87 $^0/_{00}$ and $\delta$D = 486 $^0/_{00}$ in Murchison from methanol extracts and suggested an association with amino acids. In the first comprehensive study of $^{15}$N chemistry, (Pizzarello et al. 1994) broke down the $^{15}$N components in Murchison as follows:
*Total volatile bases* (C$_1$ – C$_6$ aliphatic amines and ammonia) $\delta$ $^{15}$N = 93.4 $^0/_{00}$
*Ammonia* $\delta$ $^{15}$N = 68.5 $^0/_{00}$
*Amino acids* (Fraction III, glycine and alanine) $\delta$ $^{15}$N = 104 $^0/_{00}$
*Polar hydrocarbons* $\delta$ $^{15}$N = 102 $^0/_{00}$
The wide range of chemicals having $^{15}$N enhancement in Murchison was attributed to possible aqueous processing in the asteroidal parent body, perhaps diluting an original carrier of interstellar origin. In two CR-type meteorites from Antarctica, (Pizzarello and Holmes 2009) found a more compact array of amino acids dominated by glycine and alanine, with $^{15}$N enhancements of 125 and 131 $^0/_{00}$ in glycine. However, D enhancements, while as high as 7,245 $^0/_{00}$ for 2-Aminoisobutyric acid, did not correlate with $^{15}$N enhancements as observed in Murchison, leading to the conclusion that here possibly two epochs of chemistry had occurred, with differing isotopic input. In summary, there is evidence in both CM and CR meteorites of a diverse suite of $^{15}$N bearing molecules, and of aqueous chemistry that could have diluted one or more highly enriched $^{15}$N primary components whose nature is still unknown.

In mass spectrometry of polymer amide comprising mostly glycine in Allende and Acfer 086 (McGeoch and McGeoch 2015; 2017) noted enhanced heavy isotope satellites at mass peaks in the 800-1200 Dalton (Da) range that corresponded to identifiable amide polymers. This isotope signal provided clear proof of extra-terrestrial origin, however the $\Delta$m = +1 and $\Delta$m = +2 enhancements could have been due to D, $^{13}$C, $^{15}$N or $^{18}$O either individually or in combination. The present work therefore employs a different technique, that of focused ion beam milling and time-of-flight secondary ion mass spectrometry (FIB/TOF/SIMS), to observe the $^{15}$N carrier and explore the link between amino acid polymers and $^{15}$N enhancement that is suggested by the prior work.

FIB/SIMS has been applied in a number of prior studies (Busemann et al. 2006; Briani et al. 2009; Bonal et al. 2010 and Floss et al. 2014) to reveal $^{15}$N and D "hot-spots" in raster scans over 10- to 100- micron areas of polished specimens. In each case $^{15}$N was derived from the ratio of $^{12}C^{15}N$ to $^{12}C^{14}N$ negative ions, which is the diagnostic used in the present work. The large $^{15}N$ enhancements in "hot spots" range from 500 $^0/_{00}$ to 5,000 $^0/_{00}$, depending upon the sample.

## MATERIALS AND METHODS

**Chondritic carbonaceous meteorites as the source of fine particles for FIB/TOF isotope analysis**

The Harvard Mineralogical and Geological Museum provided meteorite samples in sealed containers delivered on the day of processing to the clean room. The use of each sample is recorded. After use they were photographed to display the etch position on the sample surface, resealed and returned to the Museum.

Details of the meteorites are as follows:

**Acfer-086** (Fig. 1A), Source: Agemour, Algeria, found 1989-90 TKW 173g, via dealer: David New 17.5g.

**Allende** (Fig. 1B), Source: Mexico 17.30g, carbonaceous (c-chondrite). Fell Feb. 8th 1969.

**Etching of meteorite samples to produce micron scale particles.**

Operatives were suited and wore powder-free nitrile rubber gloves. In a clean room extractor hood, at room temperature with high airflow, the samples were hand held while being etched to a total depth of 6mm with diamond burrs (Fig. 1C). The diamonds had been vacuum brazed at high temperature onto the stainless steel burr shafts to avoid the presence of glue of animal origin and organics in general. Etching on a cut/slice face (not an original exterior weathered face) was via slow steady rotation of a burr under light applied force via a miniature stepper motor that did not have motor brushes and did not contribute metal or lubricant contamination to the clean room. Two shapes of burr were used, the larger diameter type, in two stages, to create a pit of diameter 6mm and depth 6mm, and the smaller conical burr to etch a finely powdered sample, of approximately 1μm particles, from the bottom of the pit without contacting the sides. After each stage the pit contents were decanted by inversion and tapping the reverse side and a new burr was used that had been cleaned by ultrasonics in deionized distilled (DI) water followed by rinsing in DI water and air-drying in a clean room hood. The powder from the third etch was decanted with inversion and tapping into a glass vial. Sample weights were in the range 2-8mg. The Acfer 086 and Allende powder samples of the present work were small fractions of the same powder samples used to obtain polymer amide spectra with dominant 4641Da components (McGeoch and McGeoch 2017) after chloroform/methanol extraction. Chondrules were avoided when choosing a place to drill.

**FIB/TOF-SIMS analysis of the samples**

Measurement of isotopes in meteorite and control samples was carried out by Time of Flight Secondary Ion Mass Spectrometry (TOF-SIMS) on a XEIA3 (TESCAN) Scanning Electron Microscope equipped with a Focused Ion Beam that used high resolution orthogonal scanning TOF-SIMS (H-TOF). The meteorite samples in powdered form were placed on standard SEM aluminum holders without a carbon/metal coating. The controls were crystallized via drying a drop of aqueous solution directly on similar stubs. The TOF-SIMS measurement was performed more than 12 hours after all samples were loaded into the vacuum chamber and pumping initiated to an ultimate vacuum of <7x10$^{-4}$ Pa. The areas to be analyzed were first cleaned by FIB with an exposure of Xe$^+$ ions at 30keV and 40nA for 180s. TOF-SIMS measurement via scanning over a 100µm x 100µm area was in all cases carried out in negative ion mode under the conditions: Xe$^+$ ion energy 30keV; beam current 1000pA for meteorite, 100pA for controls; spot size 1µm. Fast SEM scanning (electron energy 5keV, current 10nA) was used for compensation of charging during analysis of the control samples. The (voluminous) output from a data collection run was analyzed off-line by TOFWERK plus TESCAN software, which allowed images to be generated of the three-dimensional spatial distribution for any given ion species.

Images in secondary electron (SE) emission were also collected. A FIB-SE image for one Allende sample is shown in Fig. 1. The fine particles had a size range from about 1µm to 10µm. The present work differs from prior work in that here we interrogate a finely milled sample that is a mixture of particles from a final drilling volume several mm across, originating at a depth of 6mm within the meteorite. In the present work we therefore lose the original spatial relationship of different atomic and molecular species, as well as crystal and inclusion morphology. Our motive for the present approach is to avoid any handling between the moment of drilling and the dusting of the powder onto the FIB stub, in order to avoid the terrestrial isotope contamination that would result from slicing and polishing.

RESULTS

**Negative ion mode data: ratio of $C_2^-$ to $CN^-$**
Negative ion spectra were recorded at two positions on two different powder samples from each of Acfer 086 and Allende, making four spectra per meteorite. Two control crystals of potassium ferrocyanide trihydrate {$K_4[Fe(CN)_6]3H_2O$} on stubs were each probed in two positions to obtain four control spectra. The dominant negative ion was $O^-$ in all the spectra (Table 1), but the focus here is on masses 24, 25, 26 and 27 that correspond to $C_2^-$, $C_2H^-$, $^{12}C^{14}N^-$ and $^{12}C^{15}N^-$. This region of the spectrum is shown in Figs. 2 and 3 for a typical meteorite case and a control.

Only the highest electron affinity molecules from Table 1 below have significant amplitudes whereas $C^-$ and $CH^-$ with relatively low affinities are weak in all the spectra, but relatively stronger in the control spectra. In the sample spectra we find a striking constancy in the $C_2$/CN and separately the $C_2H$/CN negative ion ratios, as listed in Table 2, where the control ratios are also given.

The correlation between $C_2$ and CN negative ions is illustrated in Figs. 4 and 5, for the samples and the control. In both meteorites there is a definite relationship between $C_2^-$ and $CN^-$ signals with $C_2^-$=0.5$CN^-$, whereas the control does not show this correlation. Including all three dominant negative ion species, the carbon to nitrogen ratio R(C/N) is 2.8, 2.64 and 1.36 for samples and control, respectively, whereas a perfect control would record R(C/N) = 1.

Table 1 of electron affinities reveals the basis of these spectra. Several important molecules and radicals do not have stable negative ions, in particular $H_2$, $N_2$, CO and HCN. With the exception of $O^-$ and $OH^-$ the strongest peaks in the negative ion spectra correspond to radicals with an electron affinity of 3 electron volts or greater. Production of radicals by the ion beam is discussed below.

**Instrument Calibration and Isotope Ratios**

In data collection on the Harvard meteorite samples the $CN^-$ negative ion signal was used to determine the nitrogen isotope ratio. As part of the process a control for the terrestrial nitrogen isotope ratio was run using potassium ferrocyanide trihydrate, which is a copious source of $CN^-$ ions. The $K_4[Fe(CN)_6]3H_2O$, was crystallized out of aqueous solution.

The following terrestrial standards (IAEA, Vienna 1995) were taken as reference values:

VSMOW water $\quad R_H = {}^2H/{}^1H = 155.76 \pm 0.05 \times 10^{-6}$
VSMOW water $\quad R_O = {}^{18}O/{}^{16}O = 2{,}005.20 \pm 0.45 \times 10^{-6}$
V-PDB $\quad R_C = {}^{13}C/{}^{12}C = 11{,}237.2 \times 10^{-6}$
Atmospheric Nitrogen $\quad R_N = {}^{15}N/{}^{14}N = 3{,}612 \pm 7 \times 10^{-6}$

The $^{15}N/^{14}N$ isotope ratio was found via the relative intensity of negative ion signals at m/z = 27 and m/z = 26 corresponding to $^{12}C^{15}N$ and $^{12}C^{14}N$ negative ions, respectively. This is the method used in numerous prior reports of the nitrogen isotope ratio in meteorites (Busemann et al. 2006; Briani et al. 2009; Bonal et al. 2010 and Floss et al. 2014) although it needs a correction that has not been mentioned in prior work. On top of the mass 27 peak from $^{12}C^{15}N$ is the peak from $^{13}C^{14}N$, which ought to be of the same order of magnitude, according to the carbon isotope ratio. These two contributions are not resolved at the 2700 resolution of the present data – the minimum resolving power required would be 4,272 – so in the present case a correction has to be applied that assumes knowledge of the $^{13}C$ enhancement.

For Allende, there is a bulk $^{13}C$ deficiency relative to terrestrial of -18.6 ± 1.23 $^0/_{00}$ (Pearson et al. 2006, and references therein). Prior work (Swart et al. 1982; Swart et al. 1983) has shown that there is evidence of different isotopic carbon phases in Allende, but no readings greater than +20 $^0/_{00}$ were recorded. As to the carbon isotope ratio in Acfer 086, no data appears to be available. If we are prepared to assume that for both these CV3 meteorites $^{13}C/^{12}C$ is at the terrestrial level, a correction for $^{13}C^{14}N$ in the mass 27 signal may be applied as follows:
If we let *y* = the mass 27 signal and *x* = the mass 26 signal, then the measurement gives

$$\frac{y}{x} = \frac{\left[^{12}C^{15}N\right] + \left[^{13}C^{14}N\right]}{\left[^{12}C^{14}N\right]} = R_{NS} + R_{CS}$$

where $R_{NS}$ is the sample ratio of $^{15}N/^{14}N$ and $R_{CS}$ is the sample ratio of $^{13}C/^{12}C$. If $^{13}C$ in the sample is "terrestrial" i.e. $R_{CS} = R_{CR} = 1.124 \times 10^{-2}$, where $R_{CR}$ is the carbon reference ratio, then we can find the nitrogen sample ratio $R_{NS}$ via subtraction of the carbon ratio from the measured y/x.

In the measurements on the ferrocyanide control we initially found an $R_N$ that was systematically higher than the known terrestrial value of $3.612 \times 10^{-3}$. In fact, via measurements at different mass 26 signal amplitudes we found that the measured y/x ratio increased linearly as a function of the mass 26 signal amplitude as shown in Fig. 6, with the terrestrial value of $R_N + R_C = 1.485 \times 10^{-2}$ only being obtained in the limit of very small mass26 signals. The mass 26 signal amplitude plotted in Fig. 6 is the peak value on the scale provided by the TESCAN TOF machine. The point at zero is included so as to indicate the correct terrestrial mass 27/26 ratio = $1.485 \times 10^{-2}$. A straight-line curve fitting gives:
*Ratio* = $1.485 \times 10^{-2}$ + 0.390 x (*Mass 26 amplitude*).

The fitted line goes exactly through the expected terrestrial ratio in the limit of small mass 26 peak amplitude. What the graph indicates is an under-recording of the mass 26 amplitude that increases linearly as that amplitude increases. The reason for this is not known at present, but possibilities are:
a) the amplifier gain for negative ions was not linear versus signal level;
b) the detector for negative ions was not linear, due to electron loss;
c) the extraction of negative ions was less efficient when charge in a peak increased;
d) there could have been space-charge-dependent optical losses for propagating negative ions.
It is not necessary to know the cause in order to proceed with the meteorite sample analysis using a correction factor dependent on mass 26 that is designed to give the correct terrestrial 27/26 ratio for the control regardless of mass 26 peak amplitude. After using this correction factor in the analysis of the meteorite samples, if an increment in the heavy nitrogen isotope is seen in those relative to the control, it will be a real effect. The same correction also should be applied to the mass 27 peak, but it is about 70 times less than the mass 26 peak, so effectively that correction can be neglected. As seen in Table 3 below, the correction factors range from 0.94 to 0.96 when applied to the sample data mass27/mass26 ratios.

In order to render the sample mass27/mass26 ratio correctly the crude measured ratio has to be adjusted downward by the factor *F* where the mass26 amplitude is in machine units and

$$F = \frac{1.485 \times 10^{-2}}{\left[1.485 \times 10^{-2} + 0.390 \times (mass26 amplitude)\right]} = \frac{1}{\left[1 + 26.3 \times (mass26 amplitude)\right]}$$

Table 3 lists the data from meteorite samples with labels L0 and L1 for Allende and F0 and F1 for Acfer 086, each at two positions.

In summary, samples "L" from the Allende meteorite have a (CN) $^{15}N/^{14}N$ ratio = 5.09 ± 0.8x$10^{-3}$ and samples "F" from the Acfer 086 meteorite have a (CN) $^{15}N/^{14}N$ = 7.28 ± 0.8x$10^{-3}$. These are to be compared with the terrestrial $^{15}N/^{14}N$ ratio = 3.612 ± 0.007 x $10^{-3}$.

Using the T test, the probability of Allende being the same as terrestrial is 0.06 and its "per mil" enhancement, defined above, is $\delta^{15}N$ = 410 ± 220 $^0/_{00}$. Similarly, the probability of Acfer 086 being the same as terrestrial is 0.004 and its "per mil" enhancement is $\delta^{15}N$ = 1,015 ± 220 $^0/_{00}$

DISCUSSION

**Ion beam fragmentation**
The 30keV xenon ions cause bond breakage upon impact with the samples. They also create a small, low-density plasma in the space above the sample surface. In the present work this plasma is subject to a high frequency (1GHz) pulsed extraction voltage, so it is short lived and minimal plasma chemistry is possible. Volatile molecular components will evaporate from the surface due to ion beam heating. What can be observed in the negative ion mode is the category of newly-minted fragments that have acquired, or retained, a bound electron, that is, have become a negative ion. The carbon and nitrogen content of CV3 meteorites is rather low (Pearson et al. 2006) so most of the material to be milled is a fine-grained mineral matrix. As milling proceeds the surface is completely eroded away with all the original atoms leaving either in atomic form or as small diatomic or triatomic species. To find such a clear, fixed 1:2 ratio between CC ($C_2$) and CN negative ion amplitudes (Fig. 4), regardless of meteorite, suggests that these two fragment species are released from a common carrier. A strong candidate for this carrier is the polymer amide -C-C-N-C-C-N-C-C-N- backbone of covalently bonded carbon and nitrogen, because randomly produced diatomic fragments from this chain have the CC:CN ratio of 1:2 that is observed. However, when $C_2H$ ions are included (Table 2) the C:N total rises to 2.80 (Acfer 086) and 2.64 (Allende), which needs to be compared to the ratios in the most simple amino acid residues:
Glycine 2:1, Alanine 3:1, Serine 3:1, Cysteine 3:1
More complex residues with nitrogen atoms on the side chain can also have low C/N ratios. Those at 3:1 or less are:
Asparagine 2:1, Glutamine 2.5:1, Histidine 2:1 and Lysine 3:1.
The polymer amide that has been characterized (McGeoch and McGeoch 2017) in the same samples is very simple, resembling poly-glycine with occasional alpha carbon hydroxylations and methylations and no multiple-carbon side chains. It has a C:N ratio of approximately 2.1:1. When the possible background $C_2$ and $C_2H$ in the control is taken into account, an upper bound for the measured C:N ratio is 2.6 (Table 2) assuming zero background, and a probable value could be as low as 2.3, which is very consistent with the mostly glycine polymer that has already been characterized in the samples (of each meteorite) and previously associated with heavy isotope enhancement.

**$^{15}$N enhancements**

In regard to the large $^{15}$N enhancements in this data, there is a proviso in regard to the Acfer 086 result that it depends on there being a close to terrestrial level of $^{13}$C in Acfer 086, which is likely considering its CV3 commonality with Allende. The $^{15}$N reading of $\delta^{15}N = 1,015$ $^0/_{00}$ from the CN negative ion in Acfer 086 is comparable to the highest bulk meteoritic $^{15}$N levels (Prombo and Clayton 1985, Grady and Pillinger 1990). It also is comparable to cometary $^{15}$N levels specifically in CN where an average (Arpigny et al. 2003) given for all comets is $\delta^{15}N = 873\pm70$ $^0/_{00}$.

In the mass spectra of polymer amide fragments from Allende (McGeoch et al. 2015; 2017) the $\Delta m = +1$ and $\Delta m = +2$ isotopic satellites showed enhancements compatible with a range of combined D and $^{15}$N possibilities (McGeoch et al. 2017) that was larger than nitrogen could produce by itself at +410 $^0/_{00}$. It therefore seems likely that polymer amide in Allende should also contain a significant D enhancement.

## CONCLUSIONS

In the present work a novel approach to sample analysis via simultaneous fragmentation and isotope measurement, has revealed the likely carrier of enhanced $^{15}$N in CV3 meteorites to be polymer amide. This result supports an earlier determination that polymer amide from these meteorites was isotopically enhanced. The C:N ratios observed here are consistent with the polymer structure derived from matrix assisted laser desorption mass spectroscopy of extracts from the same samples in earlier work.

This technique, which takes a powder sample and performs analysis without further processing, is anticipated to be a useful addition to the suite of extraterrestrial probes of comets, asteroids and planetary bodies, with a focus on the analysis of nitrogen bearing species and their isotopic content.

## ACKNOWLEDGEMENTS

M. W. M. and J. E. M. M. are grateful to TESCAN for donating significant machine time and experimental effort to this study. We thank Prof. Guido Guidotti for discussions and the provision of facilities and materials.

**Table 1. Negative Ion electron affinities and relative intensity in samples**

| Mass | Negative ion species | Electron Affinity (eV) | Rel. Intensity (%) Acfer 086 | Rel. Intensity (%) Allende |
|---|---|---|---|---|
| 1 | $^1H = H$ | 0.754 | 2.8 | 2.8 |
| 2 | $^2H = D$ | 0.755 | - | - |
| 6 | $^6Li$ | 0.618 | - | - |
| 7 | $^7Li$ | 0.618 | - | - |
| 12 | $^{12}C$ | 1.262 | 0.62 | 0.42 |
| 13 | $^{12}CH$ | 1.238 | 0.42 | 0.34 |
| 16 | $^{16}O$ | 1.461 | 50.1 | 48.8 |
| 17 | $^{16}OH$ | 1.828 | 17.0 | 13.9 |
| 18 | $^{18}O, ^{16}OD$ | 1.461, 1.826 | - | - |
| 19 | $^{19}F, ^{16}OH(H_2)$ | 3.401, 1.53 | 17.7 | 21.1 |
| 23 | $^{23}Na$ | 0.548 | - | - |
| 24 | $^{12}C_2$ | 3.269 | 2.8 | 2.7 |
| 25 | $^{12}C_2H$ | 2.969 | 1.84 | 2.1 |
| 26 | $^{12}C^{13}CH$ | 2.969 | - | - |
| 26 | $^{12}C^{14}N$ | 3.862 | 4.9 | 6.1 |
| 26 | $^{12}C_2H_2$ | 0.490 | - | - |
| 27 | $^{12}C^{15}N$ | 3.862 | - | - |
| 27 | $^{13}C^{14}N$ | 3.862 | - | - |
| 28 | $^{28}Si$ | 1.389 | 0.3 | 0.2 |
| 32 | $^{16}O_2$ | 0.451 | 1.54 | 1.5 |
| higher masses | | | - | - |

**Table 2. Negative ion and partial C/N ratios in samples and control**

| | Ratio (average of 4 spectra ± $\sigma/\sqrt{4}$) | | |
|---|---|---|---|
| | $C_2^-/CN^-$ | $C_2H^-/CN^-$ | $R(C/N) = \dfrac{2C_2^- + 2C_2H^- + CN^-}{CN^-}$ |
| Acfer 086 | 0.52 ± 0.05 | 0.38 ± 0.06 | 2.80 ± 0.02 |
| Allende | 0.48 ± 0.01 | 0.34 ± 0.01 | 2.64 ± 0.02 |
| $K_4[Fe(CN)_6]3H_2O$ | 0.1 ± 0.06 | 0.08 ± 0.04 | 1.36 ± 0.21 |

**Table 3. Data from samples with mass26 amplitude correction applied**

| Sample ID | mass 26 signal | Crude 27/26 ratio y/x | Correction factor $F$ | Corrected 27/26 ratio y/x | Adjusted y/x–$^{13}C/^{12}C$ ratio |
|---|---|---|---|---|---|
| L0 pos1 | 0.00234 | 0.0168 | 0.942 | 0.0158 | $4.56 \times 10^{-3}$ |
| L0 pos2 | 0.00159 | 0.0180 | 0.960 | 0.0173 | $6.06 \times 10^{-3}$ |
| L1 pos1 | 0.00148 | 0.0176 | 0.962 | 0.0169 | $5.66 \times 10^{-3}$ |
| L1 pos2 | 0.00245 | 0.0163 | 0.939 | 0.0153 | $4.06 \times 10^{-3}$ |
| | | | | average | $5.085 \times 10^{-3}$ |
| | | | | $\sigma/\sqrt{N}$ | $8.07 \times 10^{-4}$ |
| | | | | | |
| F0 pos1 | 0.00170 | 0.0208 | 0.957 | 0.0199 | $8.66 \times 10^{-3}$ |
| F0 pos2 | 0.00191 | 0.0187 | 0.952 | 0.0178 | $6.56 \times 10^{-3}$ |
| F1 pos1 | 0.00218 | 0.0196 | 0.946 | 0.0185 | $7.26 \times 10^{-3}$ |
| F1 pos2 | 0.00288 | 0.0193 | 0.929 | 0.0179 | $6.66 \times 10^{-3}$ |
| | | | | average | $7.28 \times 10^{-3}$ |
| | | | | $\sigma/\sqrt{N}$ | $8.4 \times 10^{-4}$ |

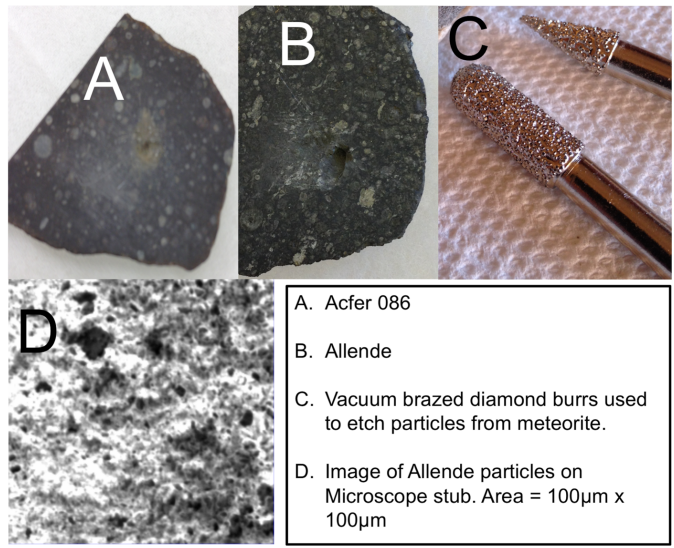

**Fig. 1. Acfer 086 (A), Allende (B) meteorite samples milled via diamond burrs (C) to produce micron scale fragments (D).**

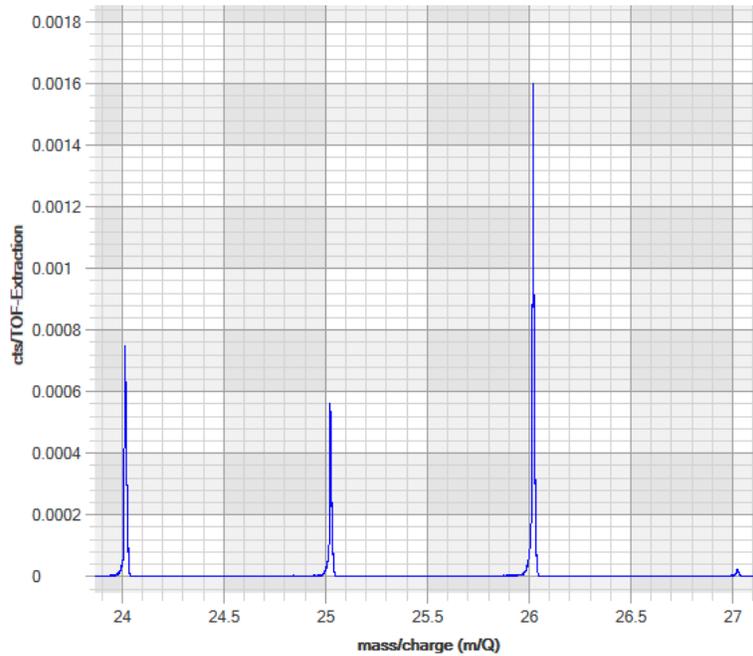

**Fig. 2. Sample negative ion spectrum (Allende) in the 24 to 27 mass range**

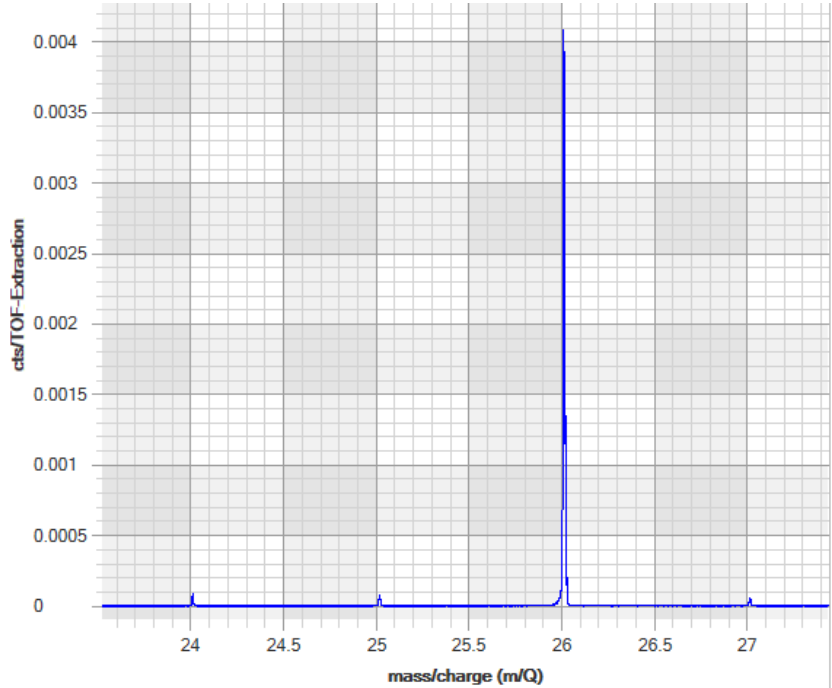

**Fig. 3. Control potassium ferrocyanide negative ion spectrum in the 24 to 27 mass range**

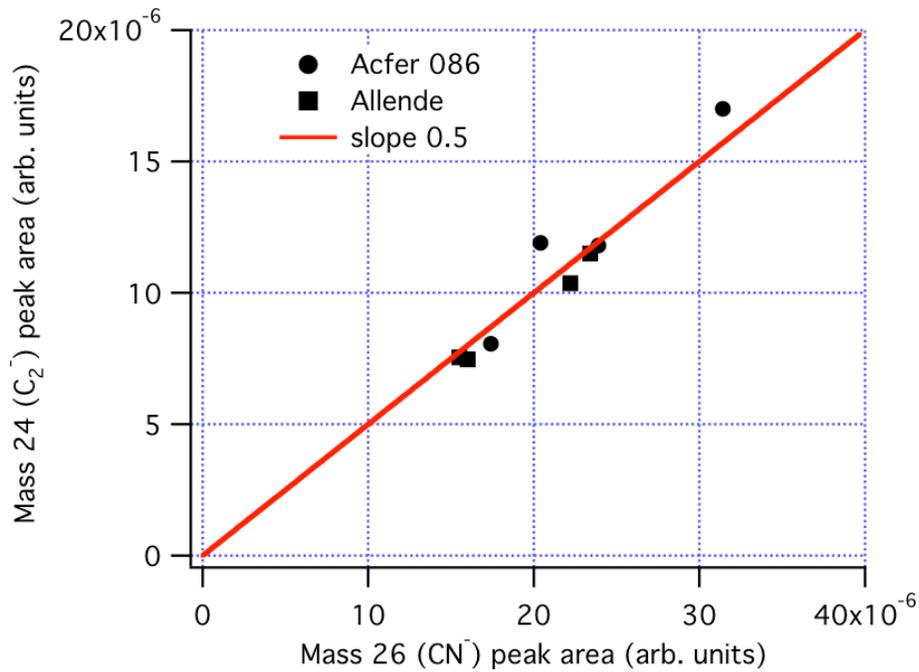

**Fig. 4. Meteorite sample $C_2$/CN trend.**

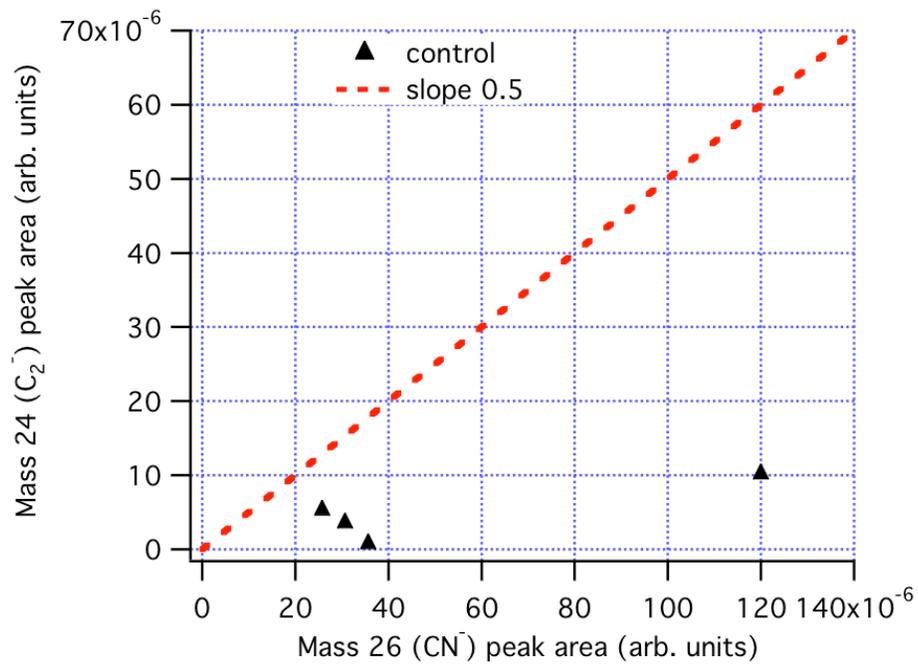

**Fig. 5. Control $C_2$/CN trend**

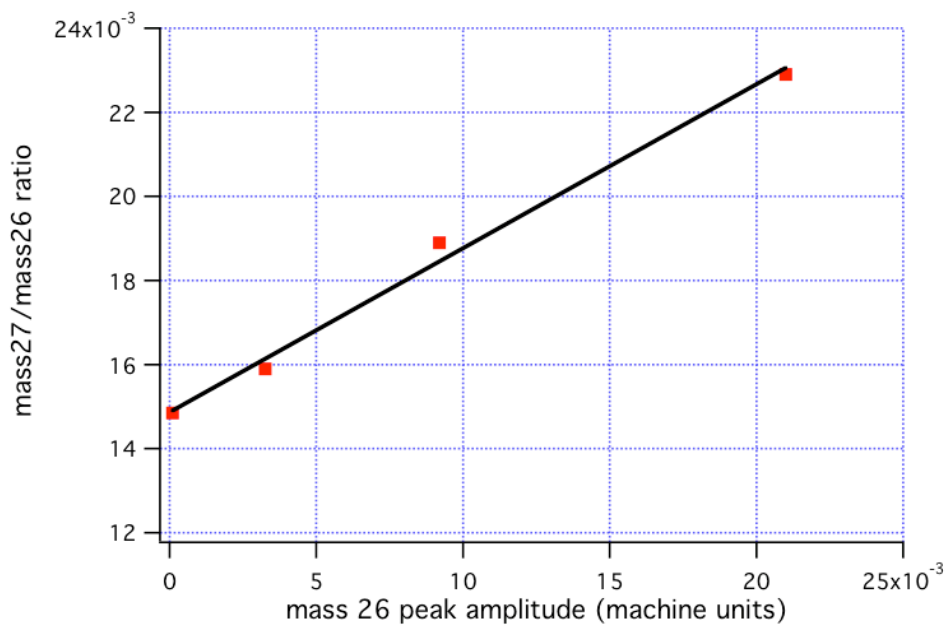

**Fig. 6. Ferrocyanide control $CN^-$ negative ion mass27/mass26 ratio vs mass 26 peak amplitude**